# Three-dimensional double helical DNA structure directly revealed from its X-ray fiber diffraction pattern by iterative phase retrieval

Tatiana Latychevskaia* and Hans-Werner Fink

*Physics Department of the University of Zurich, CH-8057 Zurich, Switzerland*
**tatiana@physik.uzh.ch*

**Abstract:** Coherent diffraction imaging (CDI) allows the retrieval of the structure of an isolated object, such as a macromolecule, from its diffraction pattern. CDI requires the fulfilment of two conditions: the imaging radiation must be coherent and the object must be isolated. We discuss that it is possible to directly retrieve the molecular structure from its diffraction pattern which was acquired neither with coherent radiation nor from an individual molecule, provided the molecule exhibits periodicity in one direction, as in the case of fiber diffraction. We demonstrate that by applying iterative phase retrieval methods to a fiber diffraction pattern, the repeating unit, that is, the molecule structure, can directly be reconstructed without any prior modeling. As an example, we recover the structure of the DNA double helix in three-dimensions from its two-dimensional X-ray fiber diffraction pattern, Photograph 51, acquired in the famous experiment by Raymond Gosling and Rosalind Franklin, at a resolution of 3.4 Å.

## 1. Introduction

In 1953 James Watson and Francis Crick deduced the structure of B-DNA from its X-ray fiber diffraction pattern which was acquired by Raymond Gosling, a PhD student of Rosalind Franklin [1-3]. Fiber diffraction [4-6] is employed for studying the molecular structure of elongated assemblies of repeating units. In the case of a subunit consisting of a few atoms, such as the DNA nucleotide, the atomic positions can be determined by combining X-ray diffraction data with the known chemical composition of the unit. In the case of larger subunits, such as proteins, the fiber diffraction pattern can provide the information about the shape of the subunit. However, to determine its detailed structure X-ray crystallography data are required. Coherent diffractive imaging (CDI) [7] is a relatively modern technique which unlike other techniques allows to directly visualize the structure of an isolated object, a macromolecule for example [7-17]. In CDI, the structure of a sample is reconstructed from its diffraction pattern by applying numerical iterative phase retrieval algorithms. For realizing CDI, two requirements must to be fulfilled: the object under study must be isolated and the imaging radiation must be coherent, although some attempts to employ partially coherent waves have also been reported [18]. The second requirement of a high brilliance coherent X-ray source is fulfilled by using a synchrotron or one of the recently build X-ray free-electron lasers. The very short X-ray pulses of the latter are to be used for imaging individual biological macromolecules [19].

In this study we show that a fiber diffraction pattern can be treated as if it was acquired in a CDI experiment. As an example, we reconstruct the three-dimensional (3D) DNA double helical structure from its two-dimensional (2D) fiber diffraction pattern by applying a phase retrieval algorithm. In this way, no modeling is thus required. The oversampling condition normally needed in the case of CDI can be relaxed when imaging a periodical structure like a DNA. Moreover, the molecule under study does not need to be isolated; the molecular structure can be directly recovered from its fiber diffraction pattern, a diffraction pattern created by an ensemble of molecules acquired with a partially coherent radiation.

## 2. Fiber diffraction and coherent diffraction imaging (CDI)

A fiber diffraction experiment, which typically employs a partially coherent X-ray source, is sketched in Fig. 1(a). A coherence length which ensures coherent addition of the waves scattered by adjacent atoms to create a peak in the diffraction pattern is sufficient. The molecules in a fiber are approximately aligned along the fiber axis, and, unlike in a crystal, they are randomly rotated along their axes. Fibers show helical symmetry [4-6]. The molecular structure is recovered in a similar to crystallographic experiment approach, starting with an initial model which is then refined to achieve a good agreement between atomic positions and intensity maxima in the fiber diffraction pattern.

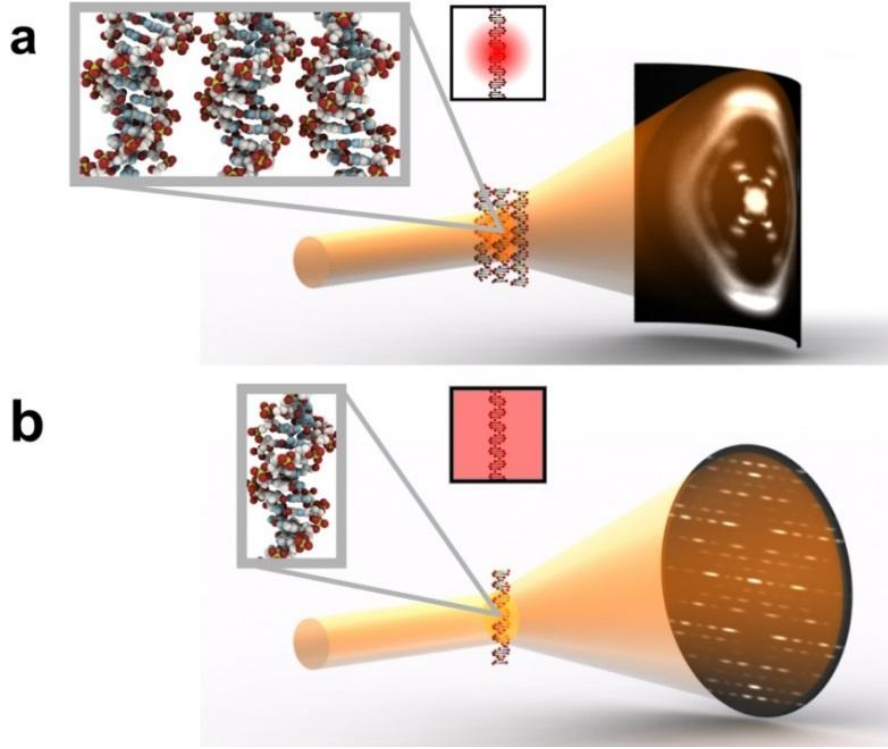

Fig. 1. Sketches of (a) fiber diffraction, as re-created from the setup of Gosling and Franklin for their DNA X-ray fiber diffraction experiment [20] and (b) coherent diffraction imaging (CDI) experiments. In a fiber diffraction experiment, the individual molecules are stretched and arranged into a fiber with a random rotation around their axis. The larger insets depict B-DNA structures rendered from the atomic positions provided by Drew et al [21]. The smaller insets illustrate the distribution of the complex coherence function (CCF) shown in red. In a fiber diffraction experiment, because of the partial coherence, the contrast of the interference pattern created by the waves scattered at different atoms within the molecule is proportional to the distance between the atoms, quickly decaying to zero as the distance between the atoms increases. In a CDI experiment, only one individual molecule is imaged at a time. The spatial coherence length in CDI is large enough to cover the entire sample area so that the waves scattered at all atoms of the molecule interfere at a degree of coherence close to 1.

A CDI experiment is sketched is in Fig. 1(b). Here, a coherent wave is employed, and the scattered wavefront in the far-field distribution is given by the Fourier transform of the distribution of atoms within the molecule. If the complex-valued distribution of the scattered wavefront could be detected, a simple inverse Fourier transform would give the molecular structure. However, since only the intensity can be recorded, the phase information is missing, but can be recovered provided the intensity distribution is sampled at a frequency finer than at least twice the Nyquist frequency (oversampled) [7]. A finer sampling of the intensity in the far-field (Fourier) domain is equivalent to zero-padding of the sample in the object domain. In reality it means that the sample must be isolated and surrounded by a non-scattering medium. The size of the sample given by $S_0 \times S_0$ should not exceed $S/2$ in each direction, where $S = 2\pi/\Delta_k$ is the extent of the reconstructed field of view and $\Delta_k$ is the pixel size in the Fourier domain. Provided this oversampling condition is fulfilled, the number of equations (that is equal to $N^2$ pixels of the sampled diffraction pattern) exceeds the number of unknowns (that is equal to $2(N/2)^2$ for a complex-valued object) and the sample structure can then uniquely be recovered from its diffraction pattern [22]. This sample structure recovery is conventionally done by applying numerical iterative methods [23, 24]. Examples of various objects, like bacteria or viruses, which were recovered by CDI from their diffraction patterns can be found in the literature [7-17].

## 3. Fiber diffraction imaging (FDI)

The method considered in the present study is the reconstruction of the repeating unit structure obtained by iterative phase retrieval of a fiber diffraction pattern. In order to distinguish the approach considered in this work from conventional fiber diffraction and CDI we name it as fiber diffraction imaging (FDI).

Given the density of an object $f(\vec{r})$ its Fourier transform $F(\vec{k})$ is given by

$$F(\vec{k}) = \int f(\vec{r}) \exp(-i\vec{k} \cdot \vec{r}) d\vec{r}, \tag{1}$$

where $\vec{r}$ is the coordinate in the object domain and $\vec{k}$ is the coordinate in the Fourier domain. For digitally sampled signals Eq. (1) can be re-written as:

$$F(p,q) = \sum_{m,n=0}^{N} f(m,n) \exp\left[-i\Delta_k \Delta_0 (mp+nq)\right], \tag{2}$$

where $\Delta_0$ and $\Delta_k$ are the pixel sizes in the object, respectively Fourier domain. When a fast Fourier transform (FFT) is applied to calculate the Fourier transform given by Eq. (2):

$$F(p,q) = \sum_{m,n=0}^{N} f(m,n) \exp\left[-\frac{2\pi i}{N}(mp+nq)\right], \tag{3}$$

and the pixel size in the object and Fourier domains are related by the following formula:

$$\Delta_k \Delta_0 = \frac{2\pi}{N}. \tag{4}$$

The measured intensity distribution is given by:

$$I(p,q) = |F(p,q)|^2 = \left|\sum_{m,n=0}^{N} f(m,n) \exp\left[-i\Delta_k \Delta_0 (mp+nq)\right]\right|^2, \tag{5}$$

which constitutes a set of equations.

Now, we consider the formation of a diffraction pattern of a sample which is an infinite one-dimensional (1D) chain of repeating units. The distribution of such a sample can be written as:

$$f(\vec{r}) = f_0(\vec{r}) \otimes \text{Ш}(\vec{r}), \tag{6}$$

where $f_0(\vec{r})$ is the transmission function of the repeating unit, $\text{Ш}(\vec{r})$ is the Shah (Dirac comb) function describing the periodical positions of the repeating units. The intensity of the diffracted wave is given by:

$$I(\vec{k}) \sim |F(\vec{k})|^2 = |F_0(\vec{k})|^2 |FT(\text{Ш})|^2 = |F_0(\vec{k})|^2 \text{Ш}(\vec{k}), \tag{7}$$

where $F_0(\vec{k})$ is the Fourier transform of the repeating unit, FT denotes the Fourier transform, and $\text{Ш}(\vec{k})$ is the Shah-function describing the positions of the diffraction peaks. According to Eq. (7) the intensity distribution of the resulting diffraction pattern is given by a product of the diffraction pattern of the repeating unit and the diffraction peaks distribution.

For a perfect infinite crystal, the signal between the diffraction peaks is zero and the intensity of the diffraction peaks is modulated by the diffraction pattern of a repeating unit $|F_0(\vec{k})|^2$ [25]. It has recently been shown that when the crystal is not perfect (as a typical realistic crystal is), then the signal between the diffraction peaks is non-zero and creates a so-called "incoherent" background which allows mapping of $\left|F_0(\vec{k})\right|^2$ and consequent reconstruction of the repeating unit $f_0(\vec{r})$ [26]. In the present study we do not consider an imperfect crystal and our method is thus not based on using the "incoherent" background between the diffraction peaks. We assume that the crystal is close to being perfect and that the intensity between the diffraction peaks is almost zero.

The principle of our approach is illustrated in Fig. 2. Figure 2(a) shows the repeating unit distribution - a cartoon of a "cat". The distribution of the "cat" occupies almost the entire region of the repeating unit cell and thus it is not oversampled within the repeating unit. The corresponding simulated pattern is shown in Fig. 2(b). Here, the diffraction pattern $I_0(\vec{k}) = \left|F_0(\vec{k})\right|^2$ was simulated by solving the analytical integral for far-field diffraction as given by Eq. (5), without applying fast Fourier transform (FFT) to avoid wrapping effects [27]. The diffraction pattern shown in Fig. 2(b) is not oversampled, it is sampled exactly at the Nyquist frequency, so that the inverse Fourier transform of the complex-valued distribution $F_0(\vec{k})$ gives the object distribution, as shown in Fig. 2(c).

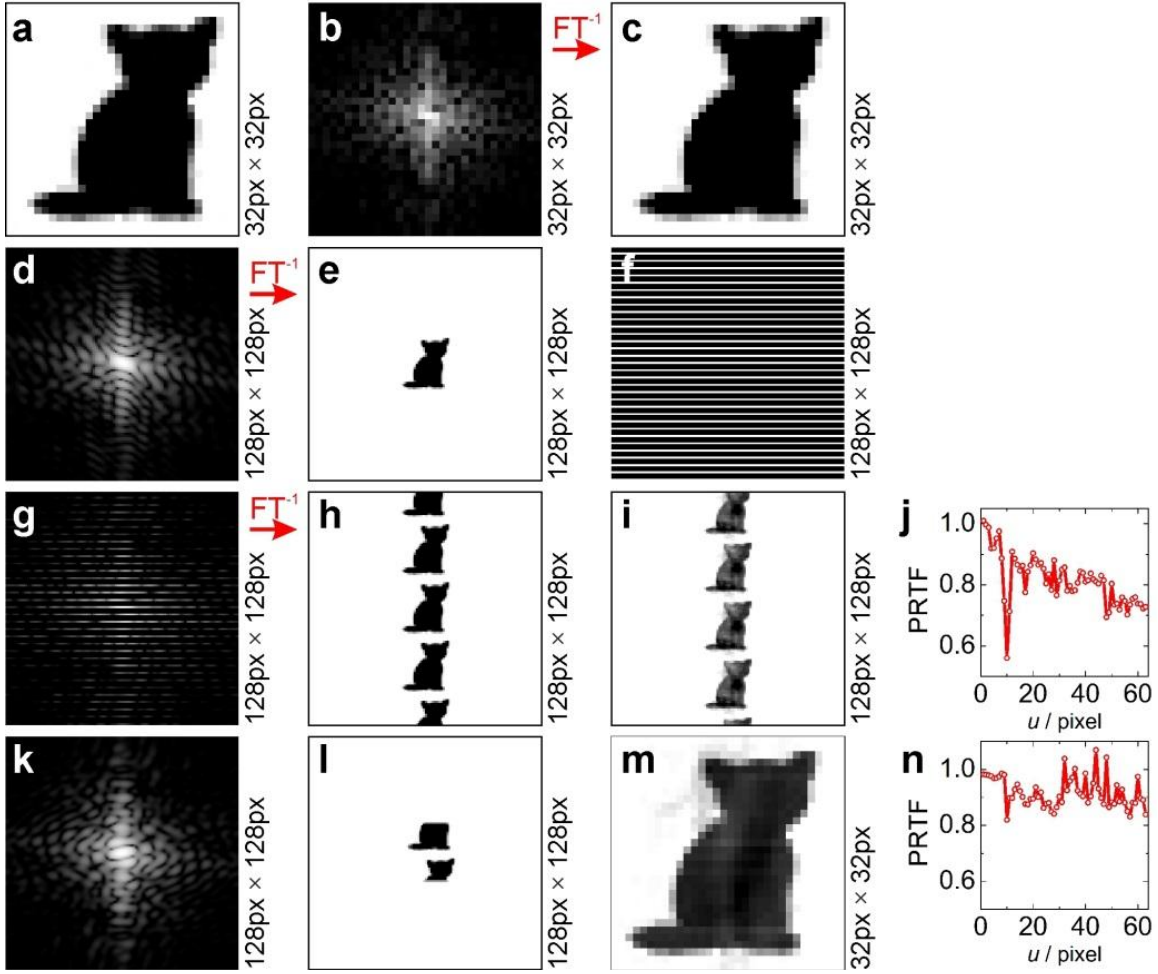

Fig. 2. Diffraction pattern of a 1D chain of repeating units. (a) Amplitude distribution of the repeating unit. (b) Diffraction pattern of the repeating unit, simulated analytically as $I_0 = |F_0(\vec{k})|^2$, not oversampled. (c) Amplitude of the inverse Fourier transform of the complex-valued distribution $F_0(\vec{k})$. (d) Diffraction pattern of the repeating unit, simulated analytically as $I_0(\vec{k}) = |F_0(\vec{k})|^2$, oversampled with $\sigma = 4$. (e) Amplitude of the inverse Fourier transform of the complex-valued distribution $F_0(\vec{k})$. (f) Distribution of the amplitude of $Ш(\vec{k})$. (g) The distribution $I(\vec{k}) \sim |F_0(\vec{k})|^2 Ш(\vec{k})$. (h) Amplitude of the inverse Fourier transform of $F_0(\vec{k}) Ш(\vec{k})$. (i) Amplitude of the reconstruction obtained from the diffraction pattern shown in (g) and the corresponding phase retrieval transfer function (PRTF) (j). (k) Diffraction pattern where the zero-intensity values were recovered to the non-zero values from the diffraction pattern shown in (g). (l) Amplitude distribution of a single reconstruction obtained from the diffraction pattern shown in (k). (m) The repeating unit distribution reconstructed from the diffraction pattern shown in (k) and the corresponding PRTF (n). The contrast is inverted in (a), (c), (e), (h), (i), (l) and (m). Diffraction patterns in (b), (d), (g) and (k) are shown in logarithmic intensity scale.

Next, a diffraction pattern $I_0(\vec{k}) = |F_0(\vec{k})|^2$ was simulated by solving the analytical integral for far-field diffraction as given by Eq. (5), where the pixel size $\Delta_k$ is four times smaller and the number of pixels $N$ is four times larger than in the previously simulated diffraction pattern, shown in Fig. 2(d). The simulated diffraction pattern is sampled at 4 times the Nyquist frequency, and thus oversampled, with an oversampling ratio $\sigma = 4$. The inverse Fourier transform of the simulated complex-valued distribution $F_0(\vec{k})$ delivers the object distribution which is zero-padded, as shown in Fig. 2(e).

Next, we create the diffraction pattern of an infinite 1D chain of repeating units, as described by Eq. (7), by multiplying the oversampled diffraction pattern of the repeating unit $|F_0(\vec{k})|^2$ with a function $Ш(\vec{k})$. The function $Ш(\vec{k})$ is given by the Fourier transform of the 1D Dirac comb-function and in two-dimensions it is a set of equidistant lines as shown in Fig. 2(f). The period of $Ш(\vec{k})$ is determined as follows. The size of the repeating unit is given by $S_0 \times S_0$ and the size of the field of view is given by $S \times S$ with $S = \frac{2\pi}{\Delta_k}$. The oversampling ratio is given by $\sigma = \frac{S}{S_0} = \frac{N}{N_0}$, where $N_0 \times N_0$ is the size of the repeating unit in pixels. The peaks due to the structure of period $S_0$ are expected at $k$-values of $k_\zeta = \frac{2\pi}{S_0}\zeta$, $\zeta = 0, \pm 1, \pm 2,..$ Or, in pixels:

$k_\zeta(\text{px}) = \frac{2\pi}{S_0 \Delta_k}\zeta = \frac{S}{S_0}\zeta = \sigma\zeta$, $\zeta = 0, \pm 1, \pm 2,..$, that is at each $\sigma^{th}$ pixel. In the simulated example considered here $\sigma = 4$ and the non-zero horizontal lines appear at every 4$^{th}$ line, as shown in Fig. 2(f). The product of $|F_0(\vec{k})|^2$ and $Ш(\vec{k})$, that is $I(\vec{k}) \propto |F_0(\vec{k})|^2 Ш(\vec{k})$, is shown in Fig. 2(g). Note that this diffraction pattern is not simply the FFT of 4 repeating units, this diffraction pattern is the diffraction pattern of an infinite 1D chain of repeating units. The inverse Fourier transform of $F_0(\vec{k}) Ш(\vec{k})$ results in the repeating units arranged into vertical line, as illustrated in Fig. 2(h).

Now, we discuss the possibility of reconstructing the repeating unit distribution from a diffraction pattern of a 1D chain of the repeating units. The intensity distribution of such diffraction pattern is given by Eq. (7). For a 2D diffraction pattern the number of equations is $N^2$. For a 2D real-valued signal, according to Friedel's law [28], the measured diffraction pattern exhibits central symmetry. Thus, the total number of equation is $N^2/2$. In the corresponding diffraction pattern, only every $\sigma^{th}$ line has non-zero values which results in a total number of lines with non-zero values: $\frac{N}{\sigma} = N_0$. Therefore, the total number of equations for a such a sample is given by $\frac{N \times N_0}{2}$. The number of unknowns corresponds to the number of pixels of the repeating unit: $N_0^2$. In order for the system of equations to have a solution, the number of equations must exceed the number of unknowns which translates into: $\frac{N \times N_0}{2} > N_0^2$ or $\sigma > 2$. When the oversampling condition $\sigma > 2$ is fulfilled the repeating unit structure can be recovered from the corresponding diffraction pattern. A typical fiber diffraction pattern has a four-quadrant symmetry, and thus the number of equations is reduced to $\frac{N \times N_0}{4}$, which leads to the oversampling condition $\sigma > 4$. We would like to emphasize that only real-valued objects (repeating unit) are considered here because it follows from the four-quadrant symmetry of a typical fiber diffraction pattern.

Similar considerations can be applied for a diffraction pattern of a 2D crystal. In this case, the number of equations corresponds to the number of measured peaks and equals to $\frac{N_0^2}{2}$ for a crystal with repeating unit of identical size in both dimensions. Since the number of unknowns is $N_0^2$ the system of equations Eq. (7), that is, the structure of the repeating unit, can thus not be solved.

The 2D diffraction pattern shown in Fig. 2(g) was reconstructed by applying iterative phase retrieval procedure employing the hybrid input-output (HIO) algorithm with a constant cylindrical object support mask (more details are provided in Appendix A). The resulting 2D reconstruction is shown in Fig. 2(i). It exhibits the repeating units arranged into a vertical line, where the distribution of the repeating unit ("cat") is correctly reconstructed. Phase retrieval transfer function (PRTF) [9, 29] was calculated by averaging 10 reconstructions with the least error (details on calculation of error are provided in Appendix A), shown in Fig. 2(j); PRTF exhibits relatively high values, decreasing to about 0.7 at the highest frequency.

Since the oversampling ratio is sufficient to recover the repeating unit, and thus the problem can have a solution, there are various ways of finding the solution. For example, the phase retrieval algorithm can be modified to reconstruct only one single repeating unit as follows: (1) In the object plane the object support mask is set to the area of $S_0 \times S_0$. (2) In the Fourier domain, the values of the pixels where the measured intensity was zero are updated with the values obtained after each iteration. This way, during the reconstruction, the zero-intensity values in the Fourier domain are recovered to non-zero values and the diffraction pattern turns into a diffraction pattern of the repeating unit, as shown in Fig. 2(k). A single reconstruction with the least error is shown in Fig. 2(l) and the repeating unit distribution obtained by averaging 10 reconstructions with the least error is shown in Fig. 2(m) exhibiting the correct recovered distribution of the "cat". PRTF was calculated by averaging 10 reconstructions with the least error (details on calculation of error are provided in Appendix A), shown in Fig. 2(n); PRTF exhibits relatively high values, decreasing to about 0.8 at the highest frequency.

An example of recovery non-binary repeating unit from the corresponding diffraction pattern is shown in Fig. 3. This example demonstrates that the method in principle should allow recovery of scatterers (such as chemical elements, for example) with different scattering amplitudes. Some artifact noise can be observed in the reconstructions shown in Fig. 2(m) and 3(d) which can be greatly reduced if a tight support mask is applied in the object domain. This can be done for example, by applying the shrinkwrap algorithm [24] which automatically adjusts the support mask tightly to the object shape.

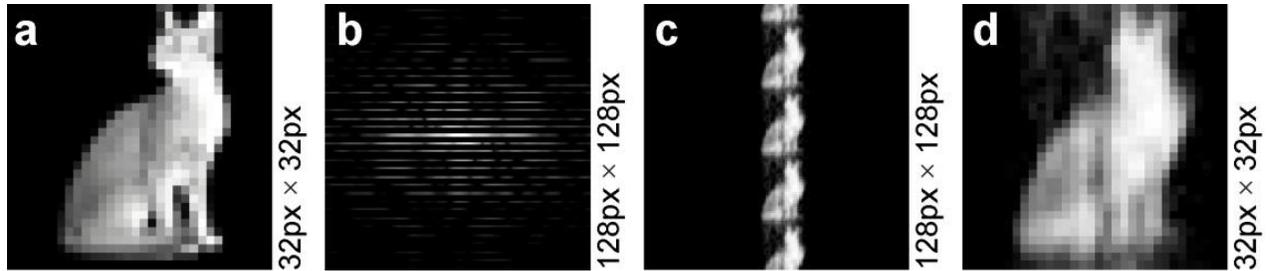

Fig. 3. Simulation and reconstruction of diffraction pattern of a 1D chain of the repeating units, non-binary repeating unit. (a) Amplitude distribution of the repeating unit. (b) The diffraction pattern calculated as $I(\vec{k}) \sim |F_0(\vec{k})|^2 \, \text{Ш}(\vec{k})$, shown in logarithmic intensity scale. (c) Amplitude of the reconstruction obtained from the diffraction pattern shown in (b) by applying the iterative phase retrieval algorithm. (d) Magnified fragment of the reconstruction shown in (c).

In all reconstructions obtained from the simulated examples and shown here, no multiple solutions were observed, the reconstructions obtained with the lowest errors all looked very similar. We can explain the uniqueness of the solution by the fact (already mentioned above) that diffraction pattern of a sample that is a 1D chain of the repeating units can be represented as a product of the oversampled diffraction pattern of the repeating unit $|F_0(\vec{k})|^2$ and a function $Ш(\vec{k})$. This means that the resulting diffraction pattern can be considered as the oversampled diffraction pattern of the repeating unit where some of measured amplitudes are missing. It has been previously demonstrated that correct sample distribution can be reconstructed from its sufficiently oversampled diffraction pattern even if only limited amount of the measured amplitudes is available (even if only 10% of the measure amplitudes are available) [30]. This situation is similar to the example shown in Figs. 2(k) – 2(m), where the zero-intensity values in the Fourier domain are recovered to non-zero values and simultaneously the repeating unit is reconstructed during the iterative procedure.

Previously, Zuo et al reported the reconstruction of a carbon nanotube where the sample was not constrained along the direction of the nanotube, which implies that the diffraction pattern was not oversampled in that direction [31]. Although Zuo et al did not discuss the uniqueness of their reconstruction, their result confirms our hypothesis that the repeating unit distribution can be reconstructed from a diffraction pattern of a 1D chain of the repeating units, provided the diffraction pattern is sufficiently oversampled to recover the repeating unit.

## 4. Simulations

Before dealing with the experimental data, we show that it is possible to reconstruct a fragment of a 3D helical molecule structure from a 3D diffraction pattern of an infinite helical molecular structure. Thereafter, we demonstrate that the 3D helical molecule structure can be reconstructed from a 2D slice of its 3D angular-averaged diffraction pattern.

### 4.1 Diffraction pattern of a single helix

We considered a sample in the form of a 3D helical curve, the helix axis is oriented along the $z$-axis. The far-field distribution of the scattered intensity is provided by the Fourier transform of the distribution of the scatterers. For a helical arrangement it has an analytical solution expressed through Bessel functions [32]:

$$U(k_x, k_y, k_z) = J_n(Kr)\exp\left[in\left(\psi + \frac{\pi}{2}\right)\right], \quad (8)$$

where $K = \sqrt{k_x^2 + k_y^2}$, $J_n(...)$ are the $n$-th order Bessel functions of the first kind, $r$ is the radius of the helix, $\tan\psi = k_y / k_x$, $i$ is the imaginary number, and $(k_x, k_y, k_z)$ are the coordinates in the Fourier domain. Amplitude of $U(k_x, k_y, k_z)$ and thus the intensity in the far-field

$$I_1(k_x, k_y, k_z) = |U(k_x, k_y, k_z)|^2 \quad (9)$$

have non-zero values only at $k_z = 2\pi n / h$ where $h$ is the period of the helix [32].

For a discontinuous helix the total intensity is given by

$$I_2(k_x, k_y, k_z) \propto \left|U\left(k_x, k_y, k_z - \frac{2\pi}{b}\right) + U(k_x, k_y, k_z) + U\left(k_x, k_y, k_z + \frac{2\pi}{b}\right)\right|^2. \quad (10)$$

Here we accounted for the zero-th (the term $U(k_x, k_y, k_z)$) and the first order (the terms $U\left(k_x, k_y, k_z - \frac{2\pi}{b}\right)$ and $U\left(k_x, k_y, k_z + \frac{2\pi}{b}\right)$) of diffraction on a discontinuous lattice, $b$ being the period of the lattice (the $z$-distance between the scatterers).

Figure 4 shows the diffraction patterns of the repeating unit and the diffraction patterns of an infinite 1D chain of the repeating units, for both continuous and discontinuous single helices. The parameters of the simulations shown in Figs. 4 – 6 were chosen to match those of an X-ray diffraction experiment on B-DNA [20]: $\Delta z = 12.75$ Å is the distance between the two helices along the $z$-axis, $h = 34$ Å is the period of the helical turn, $b = 3.4$ Å is the distance between the scatterers along the $z$-axis with 10 scatterers per helical turn. The distributions were sampled with $128 \times 128 \times 128$ pixels.

From Figs. 4(a) – 4(e) it is evident that the repeating unit in form of a one continuous helical turn provides the "X"-shaped distribution of the intensity in the diffraction pattern. When repeating units are arranged periodically, the diffraction lines (analogous to diffraction peaks for crystals) appear in the diffraction pattern, see Figs. 4(f) – 4(j). It should be noted that 3D diffraction pattern of single continuous helix according to Eq. (9) exhibits no angular dependency, which is also illustrated in the simulations shown in Fig. 4(f) – 4(j). Therefore, a single 2D diffraction pattern is sufficient for reproducing the entire 3D diffraction pattern by a simple rotation.

A discontinuous helix, made up of individual scatterers (discontinuous), with 10 scatterers per helical turn, produces a diffraction pattern which can be interpreted as a superposition of continuous helix diffraction patterns shifted along the $k_z$-axis, see Figs. 4(k) – 4(t). Here, the 3D diffraction pattern of single discontinuous helix according to Eq. (10) exhibits angular dependency, which is also illustrated in the simulations shown in Fig. 4(k) – 4(t).

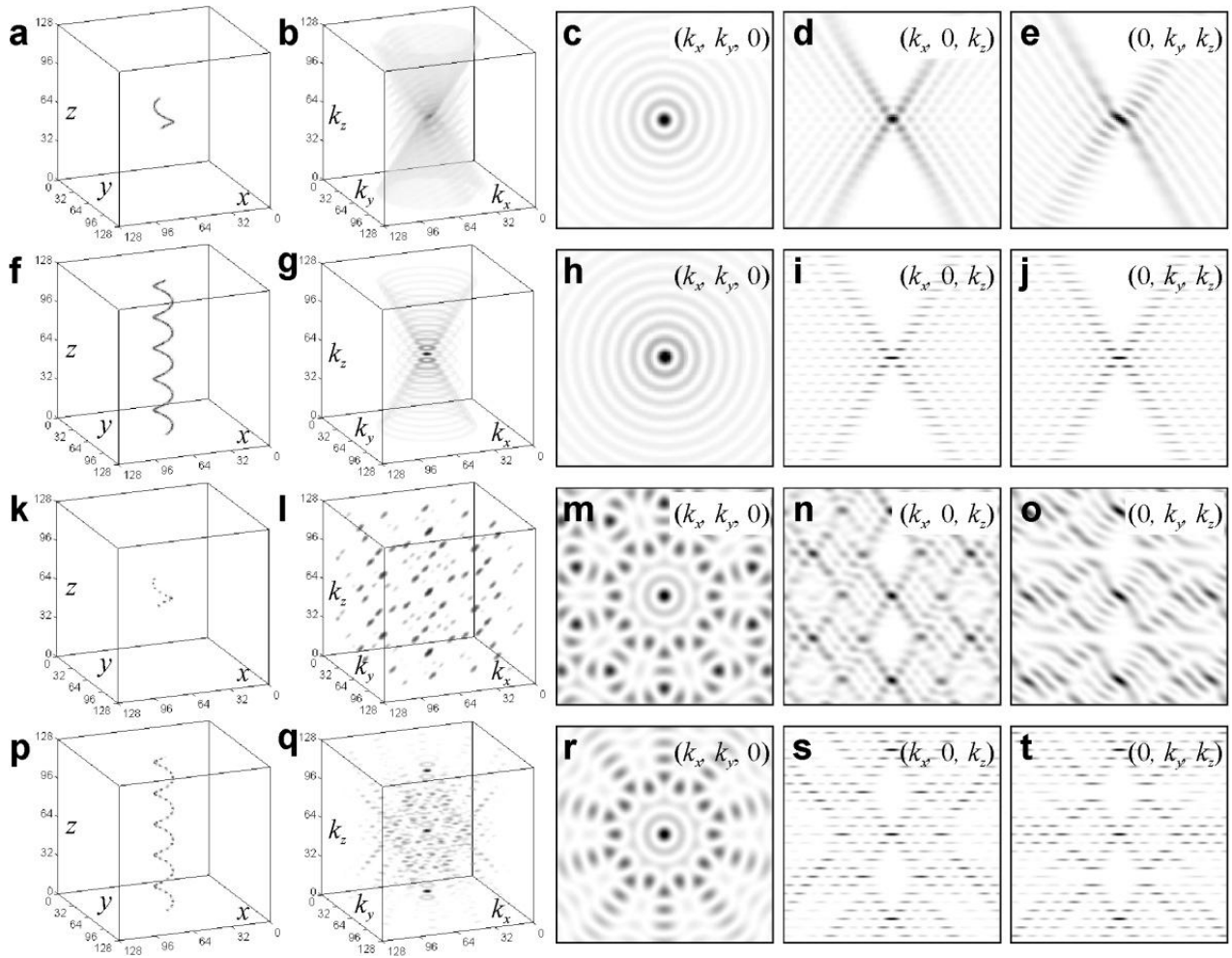

Fig. 4. Simulated diffraction patterns of single continuous and discontinuous 3D helices. (a) Amplitude distribution of the repeating unit of a continuous helix. (b) Diffraction pattern of the repeating unit shown in (a), simulated by Eq. (5). (c) – (e) 2D distributions of the intensity in the $(k_x, k_y, 0)$, $(k_x, 0, k_z)$ and $(0, k_y, k_z)$-planes, respectively. (f) Amplitude distribution of the fragment of the continuous infinite helix. (g) Diffraction pattern of the infinite continuous helix, simulated analytically by Eq. (9). (h) – (j) 2D distributions of the intensity in the $(k_x, k_y, 0)$,

$(k_x, 0, k_z)$ and $(0, k_y, k_z)$-planes, respectively. (k) Amplitude distribution of the repeating unit of a discontinuous helix with 10 scatterers per turn. (l) Diffraction pattern of the repeating unit shown in (k), simulated by Eq. (5). (m) – (o) 2D distributions of the intensity in the $(k_x, k_y, 0)$, $(k_x, 0, k_z)$ and $(0, k_y, k_z)$-planes, respectively. (p) Amplitude distribution of the fragment of the discontinuous infinite helix. (q) Diffraction pattern of the infinite discontinuous helix, simulated analytically by Eq. (10). (r) – (t) 2D distributions of the intensity in the $(k_x, k_y, 0)$, $(k_x, 0, k_z)$ and $(0, k_y, k_z)$-planes, respectively.

### 4.2 Diffraction pattern of a discontinuous double helix

For a discontinuous *double* helix the intensity distribution in the diffraction pattern is given by

$$I_3(k_x, k_y, k_z) \propto 2[1+\cos(k_z \Delta z)] \times \left| U\left(k_x, k_y, k_z - \frac{2\pi}{b}\right) + U(k_x, k_y, k_z) + U\left(k_x, k_y, k_z + \frac{2\pi}{b}\right) \right|^2, \quad (11)$$

where the modulating term $2[1+\cos(k_z \Delta z)]$ accounts for the second helix shifted by $\Delta z$ relatively to the first helix.

The 3D diffraction pattern of a discontinuous double helix simulated from the analytical solution given by Eq. (11) is shown in Fig. 5(a). The diffraction pattern exhibits a 10-fold angular dependency in the $(k_x, k_y, 0)$-plane (shown in Fig. 5(b)), as provided by the 10-fold symmetry of the arrangement of the scatterers in the $(x, y)$-plane. The 2D intensity distributions in the $(k_x, k_y, 0)$, $(k_x, 0, k_z)$ and $(0, k_y, k_z)$-planes are shown in Figs. 5(b) – 5(d), correspondingly. Since the simulated 3D object considered here is a real-valued object, according to Friedel's law, the intensity of its diffraction pattern is centrosymmetric [28]. It should be noted that the diffraction pattern simulated according to Eq. (11) corresponds to a helix not finite (but infinite) in *z*-direction.

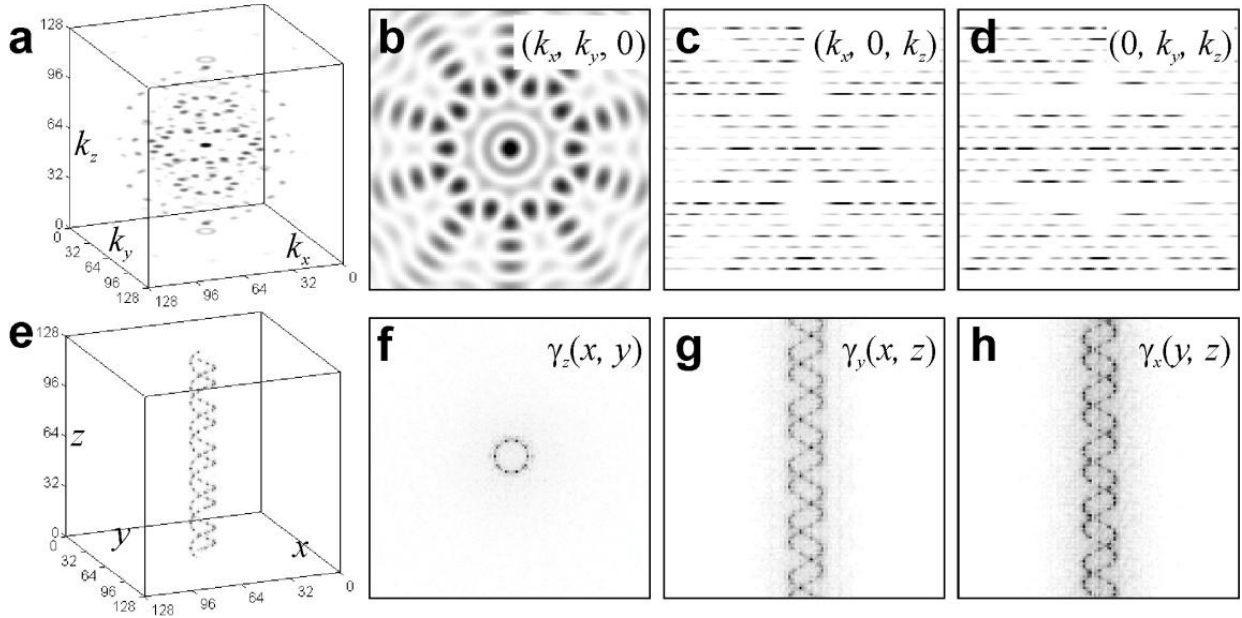

Fig. 5. Simulated 3D diffraction pattern of a discontinuous double helix and its reconstructions. (a) The simulated 3D diffraction pattern. (b) – (d) 2D distributions of the intensity in the $(k_x, k_y, 0)$, $(k_x, 0, k_z)$ and $(0, k_y, k_z)$ -planes, respectively. (e) 3D sample distribution $\gamma(x, y, z)$ reconstructed from its 3D diffraction pattern. (f) – (h) Projected distributions $\gamma_z(x, y)$, $\gamma_y(x, z)$ and $\gamma_x(y, z)$, respectively.

A 3D object reconstructed from its 3D diffraction pattern by applying an iterative phase retrieval algorithm (the reconstruction algorithm is described in Appendix A) is presented in Fig. 5(e), showing scatterers arranged into a double helix.

Projected distributions were calculated from the reconstructed 3D distribution $\gamma(x, y, z)$ as follows:

$$\gamma_x(y, z) = \int \gamma(x, y, z)\, dx,$$
$$\gamma_y(x, z) = \int \gamma(x, y, z)\, dy, \qquad (12)$$
$$\gamma_z(x, y) = \int \gamma(x, y, z)\, dz.$$

The projected distribution $\gamma_z(x, y)$ exhibits ten scatterers arranged into a circle, as shown in Fig. 5(f). The projected distributions $\gamma_y(x, z)$ and $\gamma_x(y, z)$ exhibit a discontinuous double helix, as shown in Figs. 5(g) and 5(h), respectively.

### 4.3 Angular-averaged diffraction pattern

As demonstrated in the previous subsection, a 3D diffraction pattern of a discontinuous double helix exhibits an angular dependency. However, in a fiber diffraction experiment, the individual molecules in the fiber are randomly rotated around their axes and the resulting diffraction pattern is thus an angular-averaged diffraction pattern. The intensity distribution of an angular-averaged 3D diffraction pattern is obtained by averaging the intensity distribution given by Eq. (11) over angle $\varphi$:

$$\left\langle I\left(k_{x},k_{y},k_{z}\right)\right\rangle_{\varphi} = \frac{1}{2\pi}\int I\left(k_{x},k_{y},k_{z}\right)\mathrm{d}\varphi \approx$$
$$2\left[1+\cos\left(k_{z}\Delta z\right)\right]\times\left(\left|J_{n1}\left(Kr\right)\right|^{2}+\left|J_{n2}\left(Kr\right)\right|^{2}+\left|J_{n3}\left(Kr\right)\right|^{2}\right), \quad (13)$$

where $n_1$, $n_2$ and $n_3$ are integer numbers which satisfy the following equations: $\frac{2\pi}{h}n_1 = k_z + \frac{2\pi}{b}$, $\frac{2\pi}{h}n_2 = k_z$, and $\frac{2\pi}{h}n_3 = k_z - \frac{2\pi}{b}$. An angular-averaged 3D diffraction pattern, simulated according to Eq. (13), and its slices are shown in Figs. 6(a) – 6(d). In Eq. (13), the terms which provided the angular ten-fold symmetry observed in the $(k_x, k_y, k_z = \mathrm{const})$-planes and the atomic arrangement information associated with them are gone (compare Figs. 5(b) and 6(b)). However, the terms related to the first-order diffraction ($\left|J_{n1}(Kr)\right|^2$ and $\left|J_{n3}(Kr)\right|^2$) which are also associated with the angular atomic arrangement remain. Thus, an angular-averaged 3D diffraction pattern of a helical structure described by Eq. (13) still contains information about angular arrangement of the scatterers.

A 3D object $\gamma(x,y,z)$ reconstructed from its angular-averaged diffraction pattern (the reconstruction algorithm is described in Appendix A) is shown in Fig. 6(e). Surprisingly, the projected distributions exhibit again a discontinuous double helix (Figs. 6(f) – 6(h)) just as the reconstruction obtained from non-angular-averaged diffraction pattern (Figs. 5(f) – 5(h)). Here, during the reconstruction the following observation was made: while the overall double helical shape was stable and the distances between scatterers remained constant, the position of the scatterers were "oscillating" around their averaged positions.

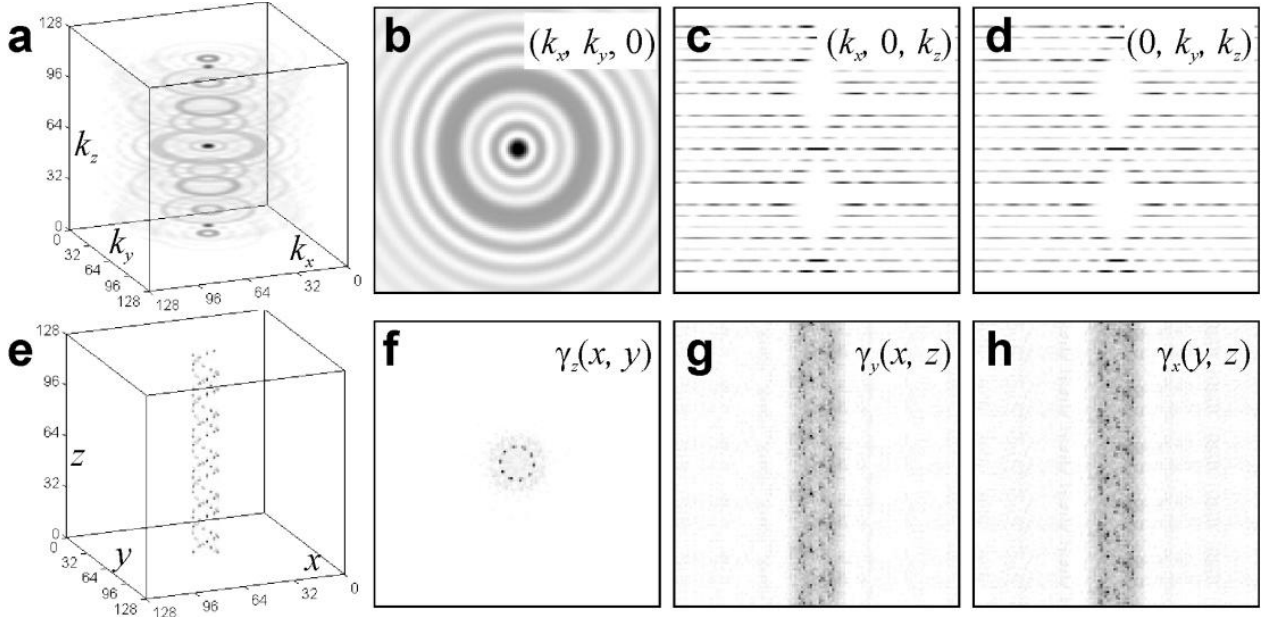

Fig. 6. Simulated angular-averaged 3D diffraction pattern of the discontinuous double helical structure. (a) The simulated 3D diffraction pattern. (b) – (d) 2D distributions of the intensity in the planes $(k_x,k_y,0)$, $(k_x,0,k_z)$ and $(0,k_y,k_z)$, respectively. (e) 3D sample distribution $\gamma(x,y,z)$ reconstructed from the angular-averaged 3D diffraction pattern. (f) – (h) Projected distributions $\gamma_z(x,y)$, $\gamma_y(x,z)$ and $\gamma_x(y,z)$, respectively.

The successful reconstruction of double helix from its angular-averaged diffraction pattern can be explained as follows. The angular-averaged 3D diffraction pattern (Fig. 6(a) – 6(d), described by Eq. 13) does not exhibit angular dependency. This is the same situation as for not angular-averaged diffraction pattern of single continuous helix (Fig. 4(f) – 4(j) described by Eq. (9)), which also does not exhibit angular dependency but 3D helical structure can be successfully recovered. The following consideration can be applied. According to the projection-slice theorem, the projected object can be reconstructed from the 2D slice through the origin of the 3D reciprocal space (for more details, see Appendix B). By rotating the slice, the angle of the projection of the object is accordingly rotated. For a helical structure, a rotation of helix around its axis is equivalent to shifting the helix along that axis. Since a lateral shift of the object only causes an additional linear phase in its Fourier spectrum, the intensity of the diffraction pattern remains unchanged when the object is laterally shifted. Thus, when the helical structure is rotated around its axis, the corresponding 2D diffraction pattern remains almost unchanged (compare Figs. 4(i) and 4(j) which are diffraction patterns of discontinuous double helix rotated by 90°).

The successful reconstruction of *individual scatterers* arranged into a double helix from its angular-averaged diffraction pattern can be explained as follows. An angular-averaged 3D diffraction pattern of a helical structure (as described by Eq. (13)) still contains information about angular arrangement of the scatterers, which can be in principle extracted in the reconstruction. When the helical structure is rotated around its axis, the corresponding 2D diffraction pattern remains almost unchanged (compare Figs. 4(i) and 4(j) for single continuous helix and Fig. 5(c) and 5(d) for discontinuous double helix). The only changes are associated with the precise positions of the scatterers. This uncertainty is within the distance between the scatterers.

A single 2D slice of the angular-averaged diffraction pattern is therefore sufficient for reproducing the entire angular averaged 3D diffraction pattern by a simple rotation. Thus, a single *2D diffraction pattern* obtained from a fiber should be sufficient for the reconstruction of the *3D structure* of the molecule (repeating unit).

*4.3 Reconstruction of other helical structures*

Figure 7 shows further examples of helical structures reconstructed from their 2D angular-averaged diffraction patterns.

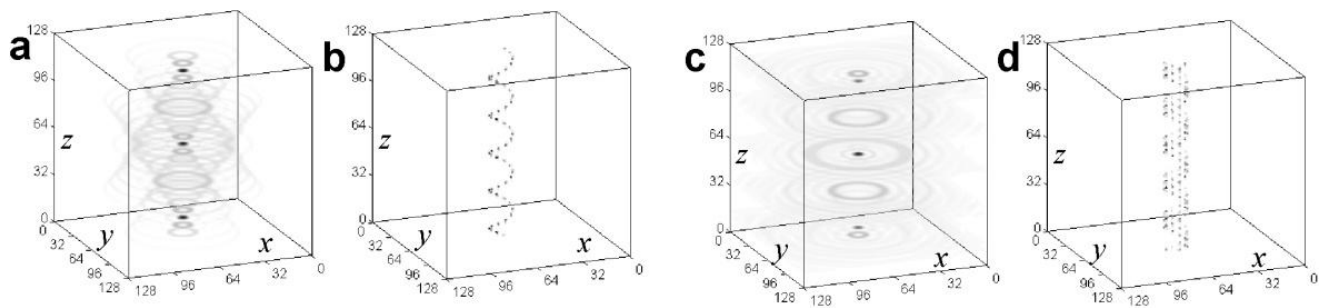

Fig. 7. Helical structures reconstructed from their 2D angular-averaged diffraction patterns. (a) Simulated angular-averaged 3D diffraction pattern of the discontinuous single helical structure and its reconstruction (b). (c) Simulated angular-averaged 3D diffraction pattern of the discontinuous triple helical structure and its reconstruction (d).

## 5. Reconstruction of the experimental X-ray DNA diffraction pattern, Photo 51

To proof the concept, we selected the fiber diffraction pattern of DNA acquired by Gosling and Franklin, which was called by the BBC as "the most important photo ever taken". The well-known dramatic story behind the Photograph 51 demonstrates how important the knowledge of the DNA molecule structure, not just its chemical composition, was. In this section we reconstruct the 3D DNA double helical shape from its 2D fiber diffraction pattern without using any prior information about the structure or any modeling.

The experimental setup, as it was employed in the X-ray fiber diffraction experiment by Gosling and Franklin, is sketched in Fig. 1(a). A fibre of calf thymus DNA, 50 μm in diameter, was placed at a distance of 15 mm in front of a photographic film which was mounted onto a cylindrical holder and kept under continuous X-ray radiation of wavelength $\lambda = 1.54$ Å for about 62 hours [20]. Parameters of Photo 51 are provided in Appendix C. A high-resolution image of the DNA diffraction pattern Photo 51 was downloaded as a scan of Raymond Gosling PhD thesis [20], it was optimized (as described in Appendix D), and is shown in Fig. 8(a). Interpretation of Photo 51 is provided in Appendix E. While the double helical structure of DNA and the periodicity of the bases can be predicted from the positions of the peaks (layer lines) in the diffraction pattern, applying a Fourier transform to the diffraction pattern does not reveal a helical structure (demonstrated in Appendix F).

By comparing the simulated and the experimental diffraction patterns, shown in Figs. 6(c) and 8(a), respectively, one immediately notices that the experimental diffraction pattern appears to be rotationally blurred. The rotational blurring can be caused by DNA molecules in the fiber being slightly rotated relatively to the axis of the fiber, and/or by bending segments of DNA where the bending can reach 20° when molecules are packed into a crystal [21], both situations are illustrated in Fig. 8(b). A good match was achieved between the experimental diffraction pattern and a sum of the simulated diffraction patterns when each simulated diffraction pattern was randomly rotated around its center with the rotations normally distributed with the standard deviation of 9° (more details are provided in Appendix G). This led to the conclusion that the DNA molecules in the fiber or their segments were randomly rotated around their axes with the angle of rotations normally distributed with the standard deviation of 9°. Next, the experimental diffraction pattern was rotationally deblurred and the result is shown in Fig. 8(c) (the deblurring procedure is described in Appendix H). It should be noted that the deblurring procedure is optional, but without deblurring only a finite fraction of the sample can be reconstructed (an example is provided in Appendix I).

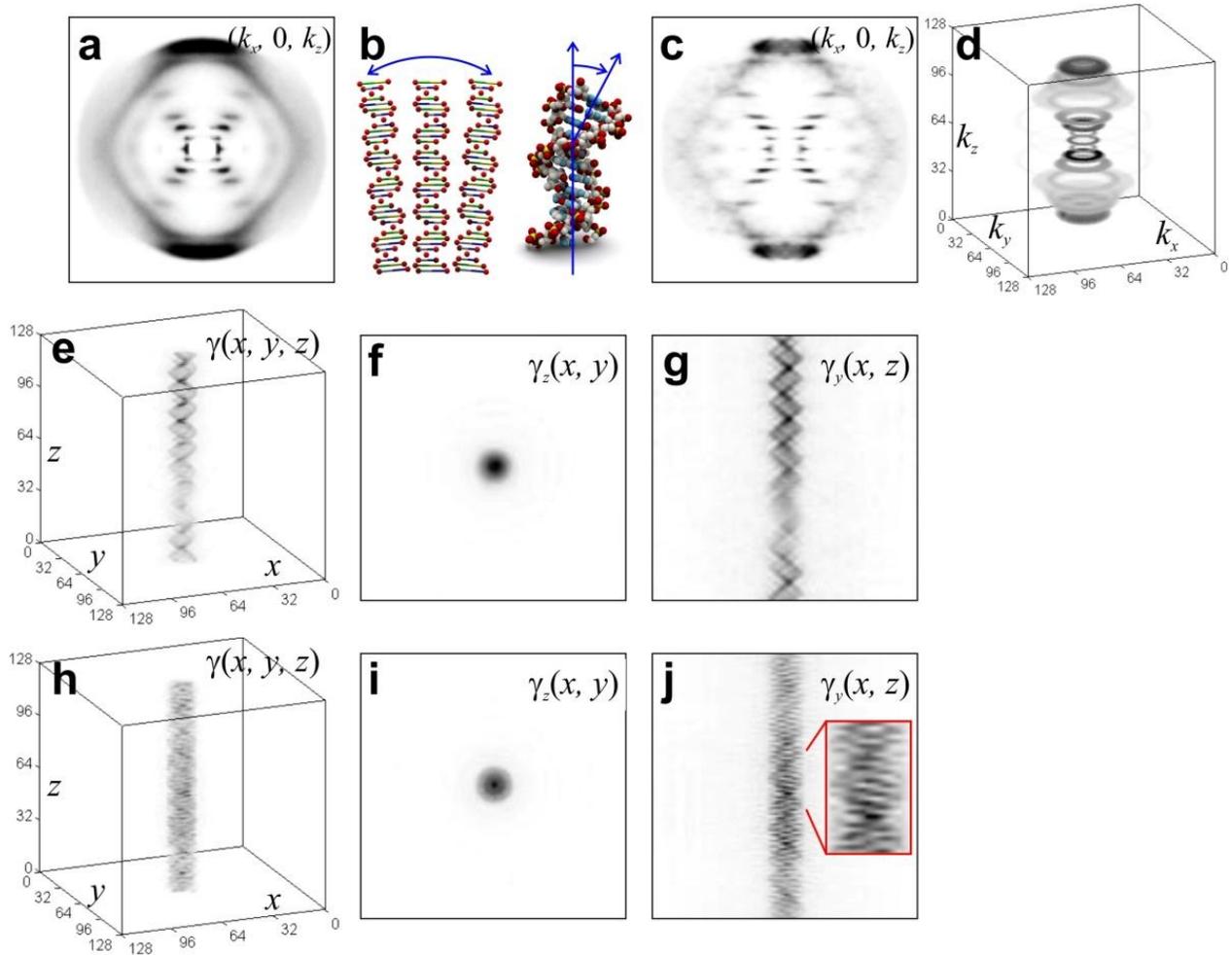

Fig. 8. Reconstruction of the experimental X-ray fiber diffraction pattern of B-DNA. (a) Experimental diffraction pattern (adapted from Refs. [3, 20]) optimized and transformed into a pattern in the $(k_x, 0, k_z)$–coordinates. (b) Illustration to possible rotation of DNA molecules within a fiber and bending of DNA segments. (c) Diffraction pattern after rotational motion deblurring. (d) 3D diffraction pattern obtained by rotation of the 2D diffraction pattern around the $k_z$-axis. (e) Low-resolution 3D distribution reconstructed from the 3D diffraction pattern where the signal corresponding to higher than 5-th order peaks (layer lines) was set to zero. (f) $\gamma_z(x, y)$ and (g) $\gamma_y(x, z)$ - projected distributions calculated from the reconstructed 3D distribution

$\gamma(x, y, z)$ as given by Eq. (12). (h) High-resolution 3D distribution reconstructed from the entire 3D diffraction pattern. (i) $\gamma_z(x, y)$ and (j) $\gamma_y(x, z)$ - projected distributions calculated from the reconstructed 3D distribution $\gamma(x, y, z)$ as given by Eq. (12).

Figure 8(d) shows the 3D diffraction pattern which was subject to the reconstruction. The low-resolution reconstructions, obtained from the 3D diffraction pattern where the signal corresponding to higher than 5-th order peaks (layer lines) was set to zero, clearly show a double helix, Figs. 8(e) and 8(g). The high-resolution reconstructions, obtained from the entire diffraction pattern, are shown in Figs. 8(h) – 8(j). The projected distributions calculated from the reconstructed 3D distribution, $\gamma_z(x, y)$, depicted in Figs. 8(f) and 8(i) show that the reconstructed double helix exhibits not a hollow but a dense center which agrees well with the molecular B-DNA model [33]. High-resolution reconstructions, shown in Fig. 8(h) and 8(j), demonstrate that the helices consist not from point-like scatterers aligned along the double helix curves, but of stick-like features connecting the opposite sides of the helix. These features can be attributed to the reconstructed bases. All the atoms that constitute a base contribute to the X-ray scattering diffraction pattern with the maxima of the scattered amplitudes for 8 keV X-rays: 1.00 (H), 6.00 (C), 7.03 (N), 8.05 (O), 15.31 (P) [34]. It is very fortunate that atoms such as P and O, that scatter the strongest due to their relatively high electron density, are positioned almost exactly on the curves that form the double helix. A closer look at the high-resolution reconstruction shows that the individual bases are not orthogonal to the helix axis but rather slightly tilted, Fig. 8(j).

## 6. Summary and conclusions

We have shown that the reconstruction of the repeating unit can be obtained from the diffraction pattern of a 1D chain of the repeating units by conventional phase retrieval methods, provided the diffraction pattern is sufficiently oversampled to recover the repeating unit. PRTFs calculated for reconstructions obtained from simulated diffraction patterns exhibit relatively high values reaching 0.7 – 0.8 at the highest frequency, thus indicating high reproducibility of the reconstructions obtained by our method.

We demonstrated our method by applying it to fiber diffraction data where typically long assemblies of repeating units are imaged. Due to the partial coherence of the sources typically applied in fiber diffraction experiments, the interference of wavefronts scattered by different molecules in the fiber can be neglected and thus the resulting diffraction pattern can consequently well be approximated by a diffraction pattern of an individual molecule (repeating unit). A fiber diffraction pattern can thus be treated as if it was recorded in a CDI experiment. For a periodical object, the CDI requirement that the sample must be isolated (zero-padded in a mathematical sense) is replaced by the much more relaxed condition that the size of the reconstructed area should exceed at least twice the size of the repeating unit. When applying our method to an angular-averaged diffraction pattern given by a fiber diffraction experiment, the 3D structure of the molecule can be reconstructed from the 2D diffraction pattern without any need for modeling or prior information about the molecule as it is required by conventional reconstruction procedure in fiber diffraction methods [4-6]. The retrieved structure is not a uniquely-defined structure of a particular isolated molecule, but is a molecular structure averaged over an ensemble, that is - structure of the repeating unit. The achievable resolution by this approach is defined by the highest resolvable peak in the diffraction pattern. For example, for the X-ray DNA diffraction pattern Photo 51, the individual bases are resolved in the reconstruction which corresponds to the resolution of 3.4 Å. While our approach recovers the geometrical arrangement of the scatterers, the contrast of a reconstructed scatterer is proportional to its scattering amplitude and thus the chemical nature of individual scatterers composing the molecule can in principle be derived.

Fiber diffraction allows imaging of some of the molecules that cannot be imaged by X-ray crystallography or CDI, where our technique can simplify the structure determination. The vast amount of already acquired fiber diffraction patterns can in principle be post-processed by using our technique and molecular structures can directly be viewed, as we have demonstrated with the historical DNA diffraction pattern.

Recently, Seuring et al demonstrated an application of the X-ray free electron laser radiation for imaging individual tobacco mosaic viruses (a helical structure) deposited onto graphene [35] where the resulting diffraction patterns are analogous to that obtained in a fiber diffraction experiment [36]. This opens up a new application for the fiber diffraction technique where the reconstruction method proposed here can serve as a simple tool for the reconstruction of the 3D molecular structure from a 2D experimental fiber diffraction pattern.

## Appendix A: Phase retrieval algorithms

### (a) Hybrid input-output algorithm

The employed hybrid input–output (HIO) algorithm [23] with the feedback parameter $\beta = 0.9$ includes the following steps. In the first iteration, random phases are combined with the measured amplitudes. The real part of the inverse Fourier transform of the resulting complex-valued distribution provides the object distribution $g_k(x,y,z)$ ($k=1$ for the first iteration). Then, the iterative procedure is applied, which includes the following steps:

(i) A Fourier transform of $g_k(x,y,z)$ provides the complex-valued distribution $G_k(k_x,k_y,k_z)$, where $(k_x,k_y,k_z)$ are the coordinates in the detector plane.

(ii) The iterated amplitudes $|G_k(k_x,k_y,k_z)|$ are replaced with the measured amplitudes $G_{\exp}(k_x,k_y,k_z)$ and thus an updated distribution $G'_k(k_x,k_y,k_z)$ is formed: $G'_k(k_x,k_y,k_z) = G_{\exp}(k_x,k_y,k_z)\exp\left(i\mathrm{Arg}\left(G_k(k_x,k_y,k_z)\right)\right)$. At the points, where the measured intensities are missing (central part of diffraction pattern), the iterated amplitudes are adapted for the next iteration: $G'_k(k_x,k_y,k_z) = |G_k(k_x,k_y,k_z)|$.

(iii) The real part of the inverse Fourier transform of $G'_k(k_x,k_y,k_z)$ provides $g'_k(x,y,z)$.

(iv) The object distribution $g_{k+1}(x,y,z)$ is calculated as follows:

$$g_{k+1}(x,y,z) = \begin{cases} g'_k(x,y,z) & \text{if } (x,y,z) \in S_1 \\ g_k(x,y,z) - \beta g'_k(x,y,z) & \text{if } (x,y,z) \notin S_1 \end{cases},$$

where $S_1$ denotes the set of the points where $g'_k(x,y,z)$ satisfies the support and the reality and positivity constraints, so that $\mathrm{Re}\{g'_k(x,y,z)\} > 0$. The resultant $g_{k+1}(x,y,z)$ is plugged into the next iteration starting at (i).

The object support mask is selected in form of a cylindrical mask with a fixed radius of 9 pixels.

The quality of the reconstruction was evaluated by calculating the error function in the object plane as it is more appropriate for the HIO algorithm [22, 23]:

$$\mathrm{Error} = \frac{\sum_{(x,y,z)\notin S} |g'_k(x,y,z)|}{\sum_{(x,y,z)\in S} |g'_k(x,y,z)|}. \tag{14}$$

For the 3D reconstructions, 1 out of 10 reconstructions with the least error was selected. Although one reconstruction with the least error is in principle sufficient, averaging over several reconstructions with the least error provides a final reconstruction with superior contrast. While we did such an averaging in the 2D case, it was however not feasible for the 3D case due to limited computation power.

### (b) Reconstruction of the experimental diffraction pattern

The reconstruction of the experimental diffraction pattern was done by applying the error-reduction [23] algorithm. The missing intensity values due to the beam stop were replaced with the iterated values after each iteration. The radius of the cylindrical object support mask in the object was varied and the reconstructions with the least error were obtained when the radius of the cylindrical object support mask was 9 pixels (corresponding to approximately 12 Å). The quality of the retrieved complex-valued amplitudes in the detector plane was evaluated by calculating the mismatch between the measured and the iterated amplitudes, or the error [23]:

$$\text{Error} = \frac{\sum_{i,j=1}^{N} \left| G_{\text{exp}}(i,j) - |G_{\text{it}}(i,j)| \right|}{\sum_{i,j=1}^{N} |G_{\text{exp}}(i,j)|}, \qquad (15)$$

where $G_{\text{exp}}(i,j)$ are the experimentally measured amplitudes at the detector, $|G_{\text{it}}(i,j)|$ are the iterated amplitudes, $i$ and $j$ are the pixel numbers $i,j = 1...N$, and the missing pixels were excluded from the summation. A total of 200 iterations were done for each reconstruction, 10 reconstructions were obtained and the reconstruction with the least error was selected.

## Appendix B: Reconstruction of slices of a diffraction pattern

The intensity of a 3D diffraction pattern of an object $\gamma(x,y,z)$ is given by:

$$I(k_x, k_y, k_z) = \left| \iiint \gamma(x,y,z) \exp\left[ -i(k_x x + k_y y + k_z z) \right] dx\,dy\,dz \right|^2, \qquad (16)$$

where $(x,y,z)$ and $(k_x, k_y, k_z)$ are the coordinates in the real and the Fourier space, respectively. According to the projection-slice theorem, each slice of a 3D diffraction pattern through its origin is a 2D diffraction pattern of the projected object. For example, the 2D intensity distribution at $k_y = 0$ is given by

$$I(k_x, 0, k_z) = \left| \iint \gamma_y(x,z) \exp\left[ -i(k_x x + k_z z) \right] dx\,dz \right|^2, \qquad (17)$$

where $\gamma_y(x,z)$ is the projection defined by Eq. (12).

2D reconstructions obtained from the slices of the 3D diffraction pattern in the $(k_x, 0, k_z)$-plane are shown in Fig. 9. The parameters of the simulation of the 3D diffraction pattern are the same mentioned in the main text: $\Delta z = 12.75$ Å is the distance between the two helices along the $z$-axis, $h = 34$ Å is the period of the helical turn, $b = 3.4$ Å is the distance between the scatterers along the $z$-axis, 10 scatterers per helical turn. The simulated 3D diffraction pattern was sampled with 128 × 128 × 128 pixels. The shown reconstructions were obtained from the slices of the original and the angular-averaged 3D diffraction patterns. The reconstructions were obtained by applying the HIO algorithm [23] with the feedback parameter $\beta = 0.9$ and a constant cylindrical object support mask. The final reconstruction was obtained by alignment [37] and averaging of 10 out of 100 reconstructions with the least error. During the alignment procedure the individual reconstructions were cyclically shifted so that the replacing values were not zeros but the values of the object distribution obtained by periodical shift of the object distribution.

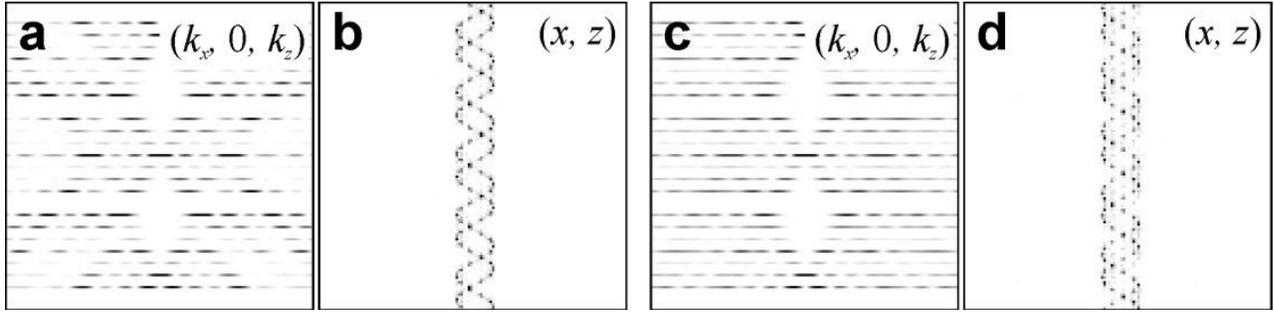

Fig. 9. 2D slices through the simulated 3D patterns and the corresponding reconstructed structures. (a) Slice through the simulated 3D diffraction pattern in the $(k_x, 0, k_z)$-plane and (b) the corresponding reconstruction of the projected object $\gamma_y(x, z)$. (c) Slice through the simulated angular-averaged 3D diffraction pattern in the $(k_x, 0, k_z)$-plane and (d) the corresponding reconstruction of the projected object $\gamma_y(x, z)$.

In both cases, the reconstructions exhibit a projected double helix where individual scatterers can be resolved. In the case of the reconstruction from the slice of the 3D diffraction pattern, every reconstruction converges to an exact solution of the projected double helix, as shown in Fig. 9(b). In the case of the reconstruction from the slice of the angular-averaged 3D diffraction pattern the projected double helix is slightly distorted. It appears as if it is viewed under a slightly different angle, which is caused by the averaging over all the angular information. Such distortions were observed for some, but not for all reconstructions. However, such distortions were not observed when reconstructing the 3D structure from the entire 3D angular-averaged diffraction pattern. Although slightly distorted, the reconstruction in Fig. 9(d) shows a double helix consisting of individual scatterers.

## Appendix C: Parameters of Photo 51

**(a) DNA sample**
The sample was a fiber of calf thymus DNA, 50 μm in diameter, which gives about $6.25 \times 10^6$ DNA molecules for a single DNA molecule of 20 Å in diameter.

**(b) Experiment**
In this historical experiment, the photographic film was mounted onto a cylindrical holder with a hole in the center to allow the direct beam to go through, which also helped to the optical alignment of the sample orientation in the beam by looking through the hole with a microscope. The distance between the sample and the photographic film was 15 mm [20] (page 66). The X-ray radiation was provided by an Ehrenberg-Spear Fine Focus tube [20] (page 52) with a copper cathode, which resulted in a wavelength of the obtained X-rays of about $\lambda = 1.54$ Å [20] (page 82).

**(c) Image of Photo 51**
The scanned page with the Photo 51 was sampled at $2475 \times 3182$ pixel with 96 dpi resolution and the effective size of the Photo 51 was about $1200 \times 1200$ pixel.

**(d) Real size of the diffraction pattern**
When a periodical structure with a period $d$ is illuminated with a coherent radiation of wavelength $\lambda$, a peak at the diffraction angle $\vartheta$ is observed in the diffraction pattern where $\vartheta$ satisfies the following equation

$$\frac{2\pi}{d} = \frac{2\pi}{\lambda} \sin \vartheta. \tag{18}$$

For $d = 3.4$ Å and wavelength $\lambda = 1.54$ Å [20] this gives $\sin \vartheta = \dfrac{\lambda}{d} = \dfrac{1.54}{3.4} = 0.45$, and therefore $\vartheta = 26.93°$.

With a distance between the sample and the photographic film of 15 mm this peak (layer line) is found at a distance 15 mm · tan $\vartheta = 15 \cdot 0.508 = 7.61$ mm from the center of the diffraction pattern. Thus, the real-world size of the X-ray fiber diffraction pattern between the top and the bottom intense peaks (layer lines) is about 16 mm × 16 mm.

**Appendix D: Optimization of Photo 51**

The diffraction pattern obtained as a scan of Photo 51, $I_0(n,m)$, was sampled with 1200 × 1200 pixels. Then it was optimized in the following way:
1. The image was centered based on the positions of the second-order diffraction peaks (layer lines), thus giving $I_1(n,m)$.
2. In general, an ideal fiber diffraction pattern exhibits a four quadrant symmetry. A real experimental fiber diffraction pattern can exhibit mirror-symmetry only with respect to the meridian due non-perfect perpendicular orientation of the fiber and the incident beam. The corresponding geometrical distortion can be corrected [38]. In the Photograph 51, the positions of the peaks (layer lines) relatively to the center were found almost symmetrical, so that no corrections for fiber tilt needed to be implemented.
2. Individual photographic film defects were numerically filtered out by setting their value to the average value of the surrounding pixels, thus giving $I_2(n,m)$.
3. The contrast of the image was inverted as $I_3(n,m) = 1/I_2(n,m)$.
4. A constant background $B_0$ was subtracted: $I_4(n,m) = I_3(n,m) - B_0$.
5. The image was rendered centrosymmetric by applying the following transformation:

$$I_5(n,m) = \frac{1}{4}\left[I_4(n,m) + I_4(n, N-m) + I_4(N-n, m) + I_4(N-n, N-m)\right], \quad (19)$$

where $N = 1200$.
6. Photo 51 was acquired on photographic film placed over a cylindrical surface, and therefore it is given in the coordinates $(L, Z)$, where $L$ is the length of the arc along the cylindrical surface, as illustrated in Fig. 10.

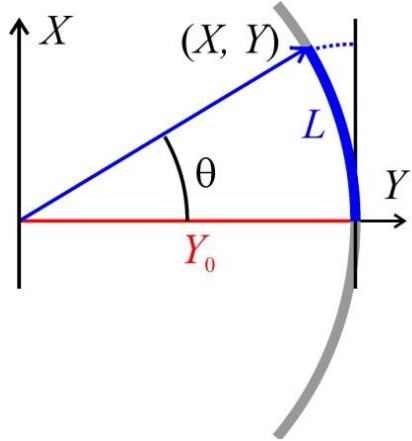

Fig. 10. Photofilm on cylindrical surface. Top view, the wavefront propagates along the $Y$-axis.

For each $(L, Z)$ coordinate, there is a corresponding $(X, Y, Z)$ coordinate on a cylindrical surface, which satisfies

$$X = Y_0 \sin\left(\frac{L}{Y_0}\right),$$
$$Y = Y_0 \cos\left(\frac{L}{Y_0}\right), \tag{20}$$
$$Z = Z,$$

where $Y_0$ is the radius of the cylinder. The $(k_x, k_y, k_z)$ coordinates are given by

$$k_x = k \frac{X}{\sqrt{X^2 + Y^2 + Z^2}} = k \frac{X}{\sqrt{Y_0^2 + Z^2}},$$
$$k_y = k \frac{Y}{\sqrt{X^2 + Y^2 + Z^2}} = k \frac{Y}{\sqrt{Y_0^2 + Z^2}}, \tag{21}$$
$$k_z = k \frac{Z}{\sqrt{X^2 + Y^2 + Z^2}} = k \frac{Z}{\sqrt{Y_0^2 + Z^2}},$$

where we used the fact that the measured intensities are on cylindrical surface so that $X^2 + Y^2 = Y_0^2$, and where $k = \frac{2\pi}{\lambda}$. Note that $k_x^2 + k_y^2 + k_z^2 = k^2$, which corresponds to the intensities measured on the Ewald sphere. The image was transformed to the $(k_x, k_z)$-coordinates as follows (similar to the procedure described in Ref. [39]).

(i) 2D array of $(k_x, k_z)$-coordinates was created.

(ii) For each $(k_x, k_z)$ coordinate, the corresponding $(L, Z)$-coordinate was evaluated from the equations:

$$X = \frac{k_x Y_0}{\sqrt{k^2 - k_z^2}}, L = Y_0 \arcsin\left(\frac{X}{Y_0}\right),$$
$$Z = \frac{k_z Y_0}{\sqrt{k^2 - k_z^2}}, \tag{22}$$

(iii) The intensity at this coordinate $(L, Z)$ was extracted and assigned to the intensity at the $(k_x, k_z)$ coordinate.

The resulting image in $(k_x, k_z)$-coordinates is $I_6(k_x, k_z)$.

7. The image was zero-padded to $1290 \times 1290$ pixels, thus giving $I_7(k_x, k_z)$.

8. The image was down-sampled to $129 \times 129$ pixels by binning with the bin size of 10, thus giving $I_8(k_x, k_z)$. The odd number of pixels ensures that the zero-order diffraction is represented by a single pixel position in the center of diffraction pattern.

9. The image was cropped to $128 \times 128$ pixels, thus giving $I_9(k_x, k_z)$.

10. After the binning and cropping, the image was double-checked to be centrosymmetric by applying the following transformation:

$$I_{10}(n,m) = \frac{1}{4}\left[I_9(n,m) + I_9(n, N-m) + I_9(N-n, m) + I_9(N-n, N-m)\right], \tag{23}$$

where $N = 128$, to ensure that the zero-order diffraction intensity is given at $N = 64$, and the intensities at pixels $N = 64 \pm \Delta N$ (for example, at pixels $(63, 63)$ and $(65, 65)$) have identical values.

**Appendix E: Interpretation of Photo 51**

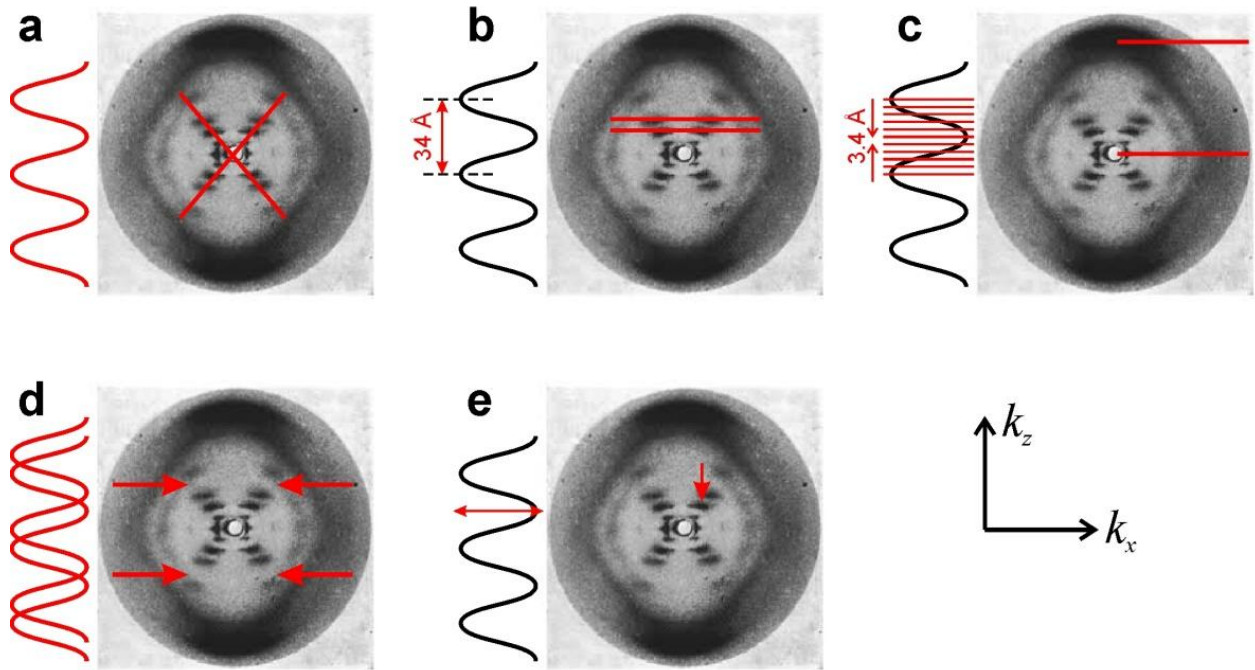

Fig. 11. Interpretation of Photo 51 B-form DNA diffraction pattern.
(a) "X"-form distribution of the diffraction peaks (layer lines) is an indication of a helical structure.

(b) The intensity exhibits non-zero values at $k_z^h = \frac{2\pi n}{h}$, where $h$ is the period of the helical turn and $n$ is the order of diffraction $n = 1, 2....$

(c) The broad extended peaks (layer lines) on the top and the bottom are formed by diffraction on small periodical features - the base pairs.
(d) The missing diffraction spots is an indication of a double helix.
(e) The position of the maxima of the diffraction spots is related to the radius of the DNA helix.

Figure 11 illustrates interpretation of Photo 51 B-form DNA diffraction pattern. The range of $k$-vectors, where $k = \frac{2\pi}{\lambda} \sin \vartheta$ and $\vartheta$ is the diffraction angle, reaches approximately $k_{max} \approx 2.18$ Å$^{-1}$. "X"-form distribution of the diffraction peaks (layer lines) is an indication of a helical structure, as shown in Fig. 11(a). As indicated in Fig. 11(b), the intensity exhibits non-zero values at $k_z^h = \frac{2\pi n}{h}$, where $h$ is the period of the helical turn and $n$ is the order of diffraction $n = 1, 2....$ The distance between the lines $\Delta k_z^h \approx 0.18$ Å$^{-1}$ corresponds to the period of the helical turn, $h = \frac{2\pi}{\Delta k_z^h} \approx 34$ Å. Figure 11(c) demonstrates that the broad extended peaks (layer lines) on the top and the bottom at $k_z^b \approx 1.8$ Å$^{-1}$ are formed by diffraction on small periodical features - the base pairs. The distance between the base pairs is therefore $b = \frac{2\pi}{k_z^b} \approx 3.4$ Å.

The missing diffraction spots, as shown in Fig. 11(d), indicate that there is a modulation of the interference pattern. The modulation is caused by interference between the waves diffracted at two helical structures that are shifted relatively to each other along their long axis by $\Delta z$ – a double helix. The diffraction pattern of a double helix can be represented as a diffraction pattern of a single helix whose intensity is modulated by $[1+\cos(k_z\Delta z)]$ function, which turns into zero at $k_z\Delta z = (2m+1)\pi$, $m=0,1,2...$ Thus, $\Delta z = \frac{(2m+1)\pi}{k_z} = \frac{(2m+1)\pi}{2\pi n}h$. In Photo 51, the missing order of diffraction is $n=4$ which gives $\Delta z = \frac{(2m+1)\pi}{8\pi}h = 4.25, 12.25, 21.25...$ Å. The shift between the two helices is approximately $\Delta z = 12.25$ Å.

The position of the maxima of the diffraction spots is related to the radius of the DNA helix, as indicated in Fig. 11(e). The intensity distribution in the individual spots in lines 0, 1, 2... are described by the Bessel functions of the first kind: $J_1(k_x r), J_2(k_x r)...$ where $r$ is the radius of the DNA. From the positions of the maxima of the diffraction spots, the radius $r$ can be recovered. For example, for the second line diffraction spots, as indicated in the figure, the intensity distribution is described by $J_2(k_x r)$ which has the first maximum at $k_x^{(2)} r = 3.05$. The position of the maximum evaluated from the diffraction pattern is about $k_x^{(2)} = 0.38$ Å, which gives $r \approx 3.05/0.38 \approx 8.0$ Å, which agrees well with the DNA radius obtained by Gosling [20].

## Appendix F: Fourier transform of Photo 51 and of the simulated 2D diffraction patterns

The Fourier transform of a diffraction pattern of an object delivers the autocorrelation of the object distribution function. Figure 12 shows the amplitudes of the Fourier transform of the optimized Photo 51 and of the simulated diffraction pattern of a discontinuous double helix. From Fig. 12 it is evident that only a limited amount of structural information can be obtained from the Fourier spectra.

For the simulated discontinuous double helix, the amplitude of the Fourier transform of its diffraction pattern, see Figs. 12(a) and 12(b), exhibit a structure which resembles two out-of phase cosines with twenty fringes per cosine period. A double helical structure cannot be established here.

The situation becomes even worse for the diffraction pattern which mimics the experimental diffraction pattern by implementing rotations of the molecules in the fiber, shown in Figs. 12(c) and 12(d). Here, the amplitude of the Fourier transform does not show any features of a discontinuous double helix, not even a periodical cosine-like structure. Similar situation is observed for Photo 51, as shown in Figs. 12(e) and 12(f).

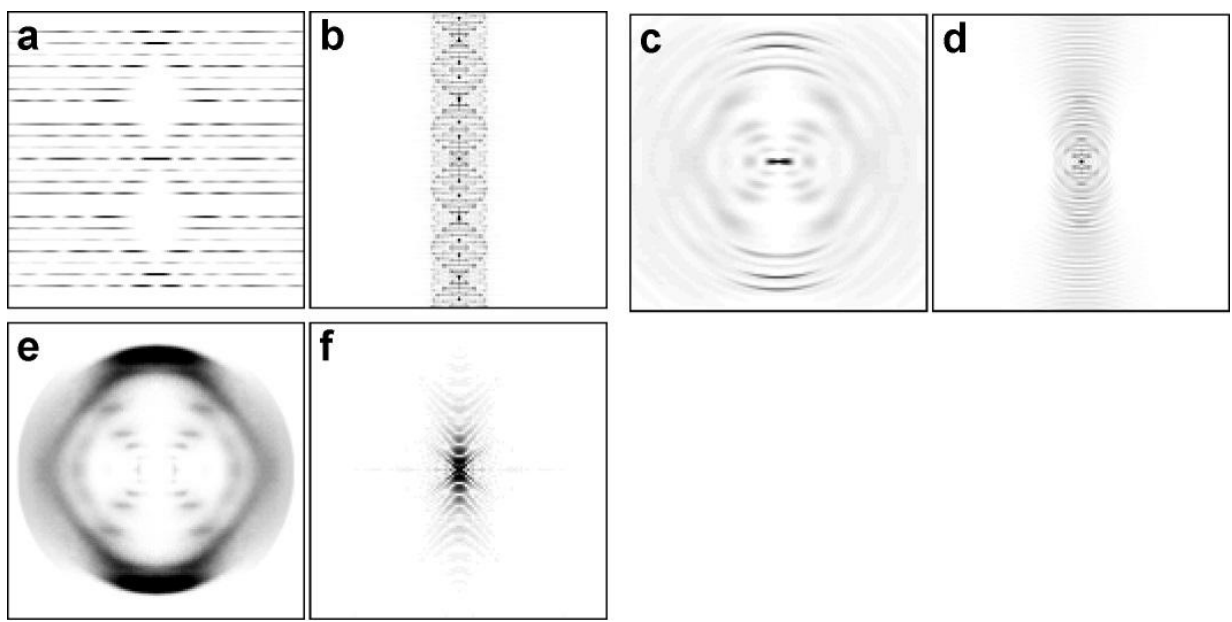

Fig. 12. Fourier spectra of the simulated 2D diffraction patterns of a discontinuous double helix and of the optimized Photo 51. (a) Simulated diffraction pattern of the discontinuous double helix and (b) amplitude of its Fourier transform. (c) Simulated diffraction pattern obtained by superposition of $10^6$ diffraction patterns of discontinuous double helices which were rotated around the direction of the incident beam with the rotation angles normally (Gaussian) distributed with the standard deviation $\sigma \approx 9°$ and (d) amplitude of its Fourier transform. (e) The optimized Photo 51 and (f) the amplitude of its Fourier transform.

**Appendix G: Effect of the rotation of individual DNA molecules within the fiber**

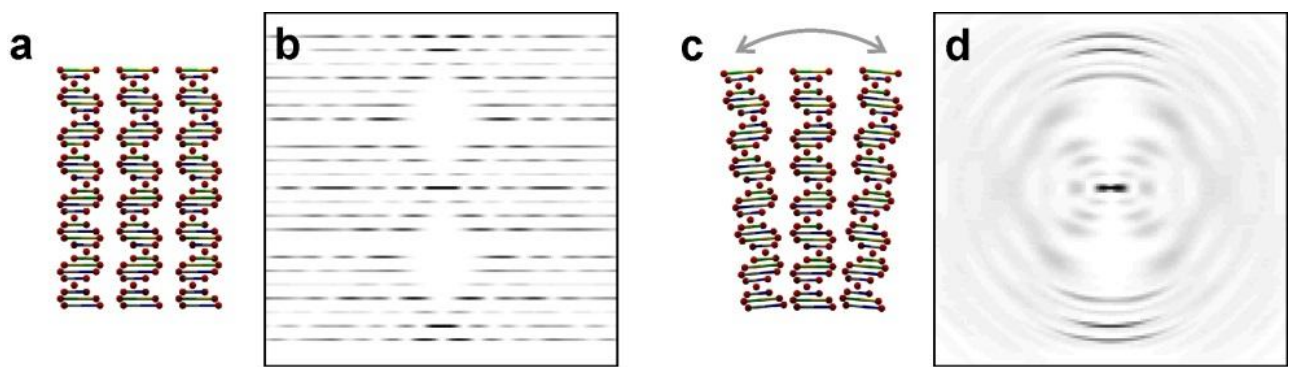

Fig. 13. Effect of rotation of individual DNA molecules in the fiber. (a) Illustration to the situation when all individual DNA molecules in the fiber are aligned and (b) the corresponding simulated diffraction pattern. (c) Illustration to the situation when some DNA molecules in the fiber are rotated and (d) the corresponding simulated rotationally blurred diffraction pattern.

The effect of rotated individual DNA molecules in the fiber is shown in Fig. 13. The diffraction pattern of a DNA fiber was simulated by incoherently adding $10^6$ diffraction patterns of individual discontinuous double helices. Each individual diffraction pattern was rotated around the direction of the incident beam and the rotation angles were Gaussian distributed with a certain the standard deviation $\sigma$. In this manner, diffraction patterns for $\sigma = 1...12°$ were simulated. Each simulated diffraction pattern was compared to the experimental diffraction pattern (optimized

Photo 51) by calculating $\text{Error} = \sum_{i,j} |I_{\exp}(i,j) - I_{\sin}(i,j)|$. The least Error was found for the simulated rotationally blurred diffraction pattern with $\sigma = 9°$, which we name RBDP$_0$. The corresponding sample arrangement and the simulated diffraction pattern RBDP$_0$ are shown in Fig. 13(c) – 13(d). Thus, the DNA molecules in the fiber imaged in Photo 51 were rotated around their axis, where the rotation angles were Gaussian distributed with a standard deviation of $\sigma = 9°$.

**Appendix H: Rotational motion deblurring**

It is assumed that the experimental diffraction pattern of DNA fiber is an incoherent sum of the diffraction patterns of rotated individual DNA. The rotation angles, $\Delta\psi$, are Gaussian distributed as $\exp\left(-\dfrac{\Delta\psi^2}{2\sigma^2}\right)$ with the standard deviation of 9°. There are currently no numerical routines that would allow for compensation of this kind of "rotational shaking" blur, although some algorithms for rotational deblurring have been presented in the literature [40]. To carry out such deblurring we notice that rotation of an image by an angle $\Delta\psi$ is equivalent to a linear shift by $\Delta\psi$ along angle coordinate $\psi$ when the image is presented in the polar $(K,\psi)$-coordinates. Thus, as a first step, the rotationally blurred experimental diffraction pattern is transformed from the spherical $(k_x, k_z)$-coordinates to the polar $(K,\psi)$-coordinates, as shown in Figs. 14(a) – 14(b). Then, at each $K$-value, 1D intensity distribution $I(\psi)$ is extracted, and it is deconvoluted with the known point spread function $\text{PSF} = \exp\left(-\dfrac{\psi^2}{2\sigma^2}\right)$, where $\sigma = 9°$. The deconvolution is performed as an iterative deconvolution according to Gold's algorithm [41], as described below. Finally, the deblurred diffraction pattern is transformed from the polar $(K,\psi)$-coordinates back to the $(k_x, k_z)$-coordinates, as shown in Figs. 14(d) – 14(e).

**Gold's iterative deconvolution algorithm**
The iterative loop includes the following steps [41]:

$$D^{(1)}(\vec{r}) = B(\vec{r})$$

(i) $\quad B^{(i)}(\vec{r}) = D^{(i)}(\vec{r}) \otimes \text{PSF}(\vec{r})$  (24)

(ii) $\quad D^{(i+1)}(\vec{r}) = D^{(i)}(\vec{r}) \dfrac{B(\vec{r})}{B^{(i)}(\vec{r})}$  (20)

(iii) $\quad i = i + 1$

Here $B(\vec{r})$ is the blurred distribution, $D^{(i)}(\vec{r})$ is the deblurred distribution at the $i$-th iteration, and $\text{PSF}(\vec{r})$ is the point-spread function. The division in step (ii) is calculated as

$$D^{(i+1)}(\vec{r}) = D^{(i)}(\vec{r}) \dfrac{B(\vec{r})\left(B^{(i)}(\vec{r})\right)^*}{\left|B^{(i)}(\vec{r})\right|^2 + \beta} \quad (25)$$

where $\beta$ is a small addend to avoid division by zero, $\beta = 10^{-6}$.

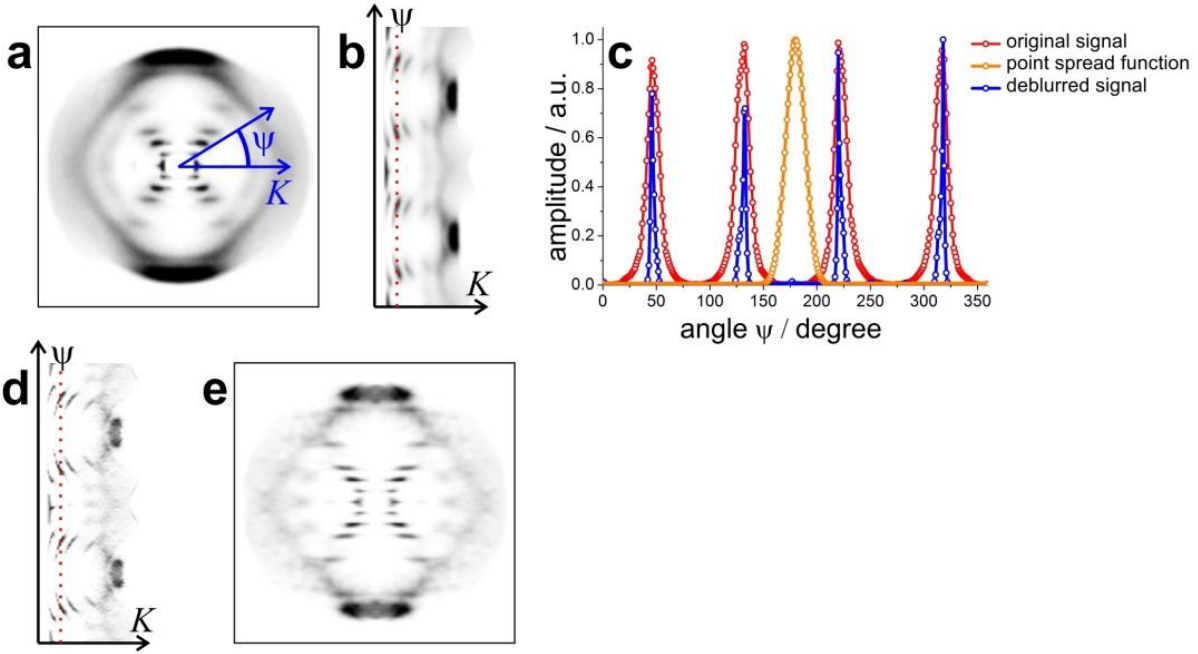

Fig. 14. Rotational motion deblurring of the experimental DNA diffraction pattern. (a) The original blurred diffraction pattern in the $(k_x, k_z)$-coordinates. (b) The same diffraction pattern in the polar $(K, \psi)$-coordinates. (c) Normalized intensity profiles of the blurred intensity, the point spread function $\mathrm{PSF} = \exp\left(-\dfrac{\psi^2}{2\sigma^2}\right)$ with $\sigma = 9°$, and the deblurred intensity obtained after the iterative deconvolution. The profiles are extracted along the red dashed line in (b) and (d), respectively. (d) The deblurred experimental diffraction pattern in the polar $(K, \psi)$-coordinates and (e) in the $(k_x, k_z)$-coordinates.

Figure 15 shows the results of the iterative deblurring applied to the simulated rotationally blurred diffraction pattern $\mathrm{RBDP}_0$. It demonstrates that the many diffraction peaks (layer lines) can be recovered although they still exhibit a somewhat blurry appearance.

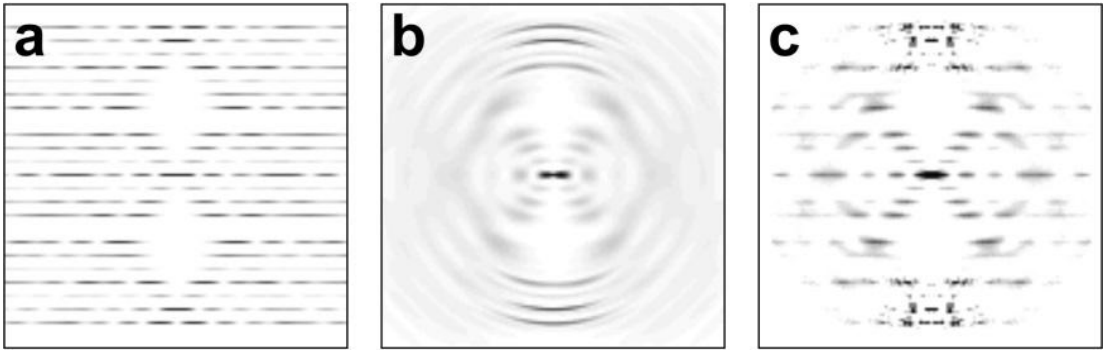

Fig. 15. Rotational motion deblurring of the simulated rotationally blurred diffraction pattern $\mathrm{RBDP}_0$. (a) The original angular-averaged diffraction pattern. (b) Simulated rotationally blurred diffraction pattern $\mathrm{RBDP}_0$. (c) The deblurred diffraction pattern obtained from (b) by the iterative deblurring.

## Appendix I: Reconstruction of the rotationally blurred angular-averaged diffraction pattern

Figure 16 demonstrates that a 3D angular-averaged diffraction pattern, obtained from a 2D rotationally blurred diffraction pattern, can in principle be reconstructed, but only a fraction of the sample can be retrieved. This can be explained as follows: a finite section of the molecule which is close to the center of rotation occupies the same position as in the same but not rotated sample.

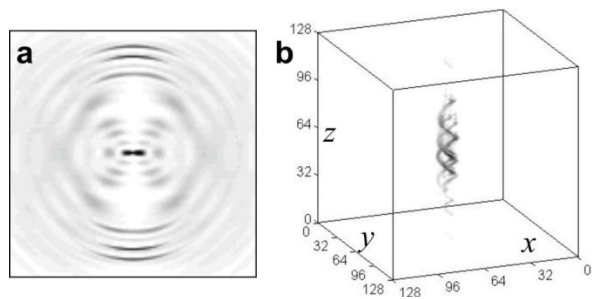

Fig. 16. Reconstruction of rotationally blurred angular-averaged diffraction pattern. (a) Simulated rotationally blurred diffraction pattern $RBDP_0$. (b) Low-resolution 3D reconstruction obtained from (a) where the signal higher than $5^{th}$ order peaks (layer lines) was set to zero.

## Acknowledgments

Financial support of the University of Zurich is acknowledged.